\documentclass[%
 reprint,
superscriptaddress,
 amsmath,amssymb,
]{revtex4-1}

\usepackage{dsfont}
\usepackage{graphicx}
\usepackage{hyperref}
\usepackage{color}

\newcommand{\im}{\mathrm{i}}

\newcommand{\drangle}{\rangle\rangle}
\newcommand{\dlangle}{\langle\langle}

\begin{document}
\title{Universal non-analytic behavior of the non-equilibrium Hall conductance in Floquet topological insulators}
%\shorttitle{Effective time reversal and echo dynamics in the transverse field Ising model} %Insert here a short version of the title if it exceeds 70 characters

%\author{Markus Schmitt\inst{1} \and Stefan Kehrein\inst{1}}
\author{Markus Schmitt}
\email{markus.schmitt@theorie.physik.uni-goettingen.de}
\affiliation{%
 Institute for Theoretical Physics, 
	Georg-August-Universit\"at G\"ottingen 
	- Friedrich-Hund-Platz 1, G\"ottingen 37077, Germany
}
\author{Pei Wang}
\affiliation{%
 Department of Physics, 
	Zhejiang Normal University 
	- Jinhua 321004, China
}
\date{\today}

\begin{abstract}
We study the Hall conductance in a Floquet topological insulator in the long time limit 
after sudden switches of the driving amplitude.
Based on a high frequency expansion of the effective Hamiltonian and the micromotion operator we
demonstrate that the Hall conductance as function of the driving amplitude follows universal 
non-analytic laws close to phase transitions that are related to conic gap closing
points, namely a logarithmic divergence for gapped initial states
and jumps of a definite height for gapless initial states.
This constitutes a generalization of the results known for the static systems to the driven case.
\end{abstract}

\maketitle

\section{Introduction}
Since the experimental discovery and theoretical explanation of the quantum Hall effect 
\cite{vklitzing,tknn} 
the concept of topological order has gained great importance in condensed matter physics 
for the understanding of phase transitions that cannot be associated with symmetry breaking.
The astonishingly robust integer quantization of the Hall conductance in units of the conductance 
quantum, $\sigma_{xy}=Ce^2/h$, is due to the fact that $C\in\mathbb Z$ 
can be identified as a topological invariant
of the underlying band structure, namely the Chern number.
After Haldane's seminal proposal of a model system featuring a quantized 
Hall conductance in the absence of an external magnetic field \cite{haldane} 
enormous experimental and theoretical efforts led to the discovery of a large variety of systems 
with similar
topologically protected transport properties, which are today referred to as topological insulators (TIs)
\cite{kane}.

Following theoretical proposals \cite{oka,kitagawa} a topological insulator 
was recently realized experimentally with ultracold fermions in a periodically shaken 
optical lattice \cite{shaken_lattice}.
Despite the absence of energy conservation such \emph{Floquet} topological insulators (FTIs) can be
characterized by the Chern number of an effective Hamiltonian and support edge modes \cite{Kitagawa1}.
This allows to tune the topological properties of the
system by adjusting the external driving force and opens possibilities to investigate non-equilibrium
signatures of topological insulators, which gained increasing theoretical attention lately 
\cite{dAlessio,Dehghani2015,Budich,Hu,Caio}. 
Note, however, that in some aspects the behavior of FTIs can significantly 
differ from the known behavior of TIs, e.g., when considering the bulk-edge correspondence 
\cite{Rudner}.

\begin{figure}[!h]
\includegraphics{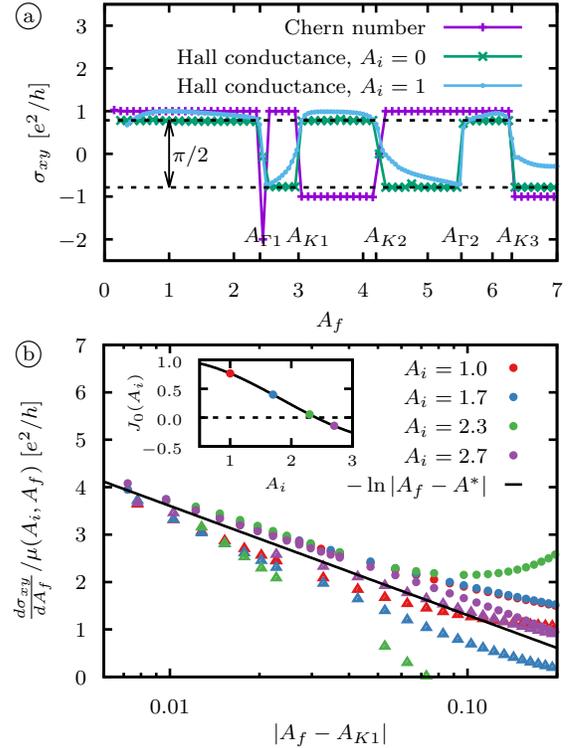}
\caption{a -- Floquet Chern number $C$ and non-equilibrium Hall conductance $\sigma_{xy}$
for quenches with $A_i=0$ and $A_i=1$ as function of $A_f$ for driving frequency $\omega=10$. 
b -- Derivative of the non-equilibrium Hall conductance
rescaled by the prefactor $\mu(A_i,A_f)=J_0(A_f)|m_i|/J_0(A_i)m_f'$ determined in 
eq. \eqref{eq:gapped_scaling} close to the transition at $A_{K_1}$. 
As $A_f$ approaches $A_{K1}$ the slopes agree increasingly well with the 
predicted $\ln|A_f-A^*|$ (black line) in
all cases even if the value $J_0(A_i)$ shown in the inset is small.
Circles/triangles denote points to the left/right of $K_1$.}
\label{fig:scaling_gapped}
\end{figure}

A situation that was studied recently by Dehghani et al. \cite{Dehghani2015} 
is the measurement of the Hall conductance
a long time after suddenly switching on the external driving force. The system is initially prepared in the
ground state of the undriven Hamiltonian $H_0$. Then the driving is suddenly switched on at time $t=0$
and 
for $t>0$ the system evolves under a time-periodic Hamiltonian $H_A(t)$, where $A$ is the driving
amplitude. The Hall conductance in the
limit $t\to\infty$ is finally obtained using linear response theory and a dephasing argument. As a result
they numerically find for an electronic system that the post-quench Hall conductance,
which is not any more an integer
multiple of the conductance quantum, exhibits sudden changes whenever the post-quench 
Hamiltonian $H_A(t)$ crosses a topological phase boundary as function of the driving amplitude $A$.
This behavior is very similar to the behavior of closed TIs after a quench, which exhibit a universal
non-analytic behavior at the ground state transition of the final Hamiltonian as shown in Refs. 
\cite{Pei1,Pei2}.

In this work we extend the analysis of closed TIs given in Refs. \cite{Pei1,Pei2} to FTIs.
We analytically investigate the behavior of the Hall conductance after sudden 
switches of the
driving amplitude for a tight binding Hamiltonian with a time periodic external potential.
The analysis is based on
high frequency expansions of the effective Hamiltonian and the micromotion operator. We focus on
phase transitions that are associated with a closing of the quasi-energy gap at the $K$-points in the
Brillouin zone. These already appear when only the first order contribution in the high 
frequency expansion of the effective Hamiltonian is considered. 
We find that suddenly switching on the driving 
amplitude from $A_i=0$ to $A_f\neq0$ with a gapless initial Hamiltonian leads to jumps of 
the Hall conductance 
by multiples of $\frac{\pi e^2}{2h}$ whenever $A_f$ crosses a phase boundary, 
which agrees with the numerical results of Ref. \cite{Dehghani2015} reproduced in 
Fig. \ref{fig:scaling_gapped}a. 
If, instead, the system is initially prepared in a quasi-stationary
Floquet mode of the initial Hamiltonian $H_{A_i}(t)$ before suddenly switching the driving amplitude
to $A_f$ the Hall conductance is continuous at critical values of $A$.
Nevertheless, it is non-analytic with a logarithmically diverging derivative 
as a function of the driving amplitude $A_f$ as shown in Fig. \ref{fig:scaling_gapped}b.

A distinct feature of FTIs is the possible presence of so-called $\pi$-edge modes \cite{Kitagawa2010,Jiang2011,Kitagawa2012,Rudner2013}. 
Note that our results do not apply to gap closings that affect these edge modes as
discussed in section \ref{subsec:universality}.

The rest of the paper is divided into two parts. In section \ref{sec:background} we introduce the
model system under consideration and briefly summarize the methods used and previous results,
which are relevant for the further analysis. In section \ref{sec:results} we present our analysis resulting
in the identification of the abovementioned non-analytic behavior of the Hall conductance, which is
universal for conic gap-closing points in two-band FTIs.

\section{Background}\label{sec:background}
In this section we introduce the model Hamiltonian under consideration and briefly review the
Floquet formalism and the high frequency expansion used for our analysis. Moreover, we give a short
summary of previous results on the non-equilibrium Hall conductance relevant for this work.
\subsection{Model Hamiltonian}
We consider a simple model Hamiltonian, namely a tight binding model on a hexagonal lattice
subject to a time-periodic external potential,
\begin{align}
	\tilde H(t)=-t_h\sum_{\langle i,j\rangle}(c_i^\dagger c_j+h.c.)+\sum_i V(\vec r_i,t)c_i^\dagger c_i\ ,
\end{align}
where 
$V(\vec r, t)=V_0\vec r\cdot\left[-\cos(\omega t)\hat e_x+\sin(\omega t)\hat e_y\right]$ 
and $\langle i,j\rangle$
denotes the set of pairs of neighboring lattice sites. This Hamiltonian
constitutes a simple description of graphene illuminated by a circularly polarized laser 
\cite{driven_graphene_experiment} or
ultracold atoms in a circularly shaken optical lattice \cite{shaken_lattice}.

\begin{figure}[t]
\includegraphics[width=.25\textwidth]{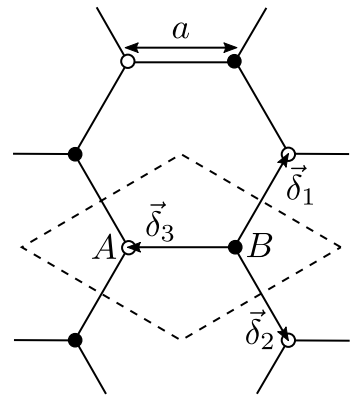}
\caption{We consider a hexagonal lattice structure. The dashed line marks a possible choice
of the unit cell with two basis sites $A$ and $B$. Depicted is moreover the unit of distance, $a$,
and the nearest-neighbor vectors $\vec\delta_i$.}
\label{fig:hex_lattice}
\end{figure}

A time-dependent gauge transformation restores translational invariance and allows to
write the Hamiltonian in momentum space as
\begin{align}
	H(t)=\sum_{\vec k} {\vec c_{\vec k}}^{\ \dagger}
	\left[\vec d_{\vec k}(t)\cdot\vec\sigma\right]
	\vec c_{\vec k}
	\label{eq:Hamiltonian}
\end{align}
(cf. appendix \ref{app:gauge-trafo}). In this expression for the Hamiltonian we introduced
\begin{align}
	\vec c_{\vec k}=\begin{pmatrix}c_{\vec kA}\\c_{\vec kB}\end{pmatrix}
\end{align}
and the coefficient vector $\vec d_{\vec k}(t)=\left(d_{\vec kx}(t),d_{\vec ky}(t),d_{\vec kz}(t)\right)^T$ 
with
\begin{align}
	d_{\vec kx}(t)&=-t_h\sum_{j=1}^3
	\cos\left((\vec k-\vec A(t))\cdot\vec\delta_j\right)
	\label{eq:dcoeff1}\ ,\\
	d_{\vec ky}(t)&=-t_h\sum_{j=1}^3
	\sin\left((\vec k-\vec A(t))\cdot\vec\delta_j\right)\ ,\\
	d_{\vec kz}(t)&=0
	\label{eq:dcoeff3}
\end{align}
as well as the vector of Pauli matrices $\vec\sigma=(\sigma^x,\sigma^y,\sigma^z)^T$.
The vectors 
\begin{align}
	\vec\delta_1=\frac{a}{2}\begin{pmatrix}1\\\sqrt{3}\end{pmatrix}\ ,\ 
	\vec\delta_2=\frac{a}{2}\begin{pmatrix}1\\-\sqrt{3}\end{pmatrix}\ ,\ 
	\vec\delta_3=a\begin{pmatrix}-1\\0\end{pmatrix}
\end{align}
are given by the differences of the positions of neighboring lattice sites (cf. Fig. \ref{fig:hex_lattice}).
Moreover,
\begin{align}
	\vec A(t)=\frac{V_0a}{\omega}\begin{pmatrix}\sin(\omega t)\\\cos(\omega t)
	\end{pmatrix}\ ,
\end{align}
where $a$ denotes the lattice spacing. In the following we will use the dimensionless driving
amplitude $A=|\vec A(t)|=V_0a\omega^{-1}$ to quantify the driving strength and, moreover, 
set $a\equiv 1$.

\subsection{Periodic driving and Floquet formalism}
In this section we recapitulate the Floquet formalism for the treatment of time-periodic Hamiltonians
and thereby introduce the notation for the subsequent discussion. We closely follow the
presentation and notation of Refs. \cite{Eckardt, Bukov}.

\subsubsection{Effective Hamiltonian and micromotion operator}
For a time-periodic Hamiltonian $H(t+T)=H(t)$ acting on a Hilbert space $\mathcal H$ 
the Floquet theorem states that the
Schr\"odinger equation
\begin{align}
	\im\frac{d}{dt}|\psi(t)\rangle=H(t)|\psi(t)\rangle
	\label{eq:seq}
\end{align}
is solved by \emph{Floquet states} of the form
\begin{align}
	|\psi_n(t)\rangle=e^{-\im\epsilon_nt}|\phi_n(t)\rangle
	\label{eq:floquet_states}
\end{align}
with \emph{quasi-energies} $\epsilon_n$ and periodic \emph{Floquet modes}
$|\phi_n(t+T)\rangle=|\phi_n(t)\rangle$ \cite{Floquet,Shirley}. Note that $\epsilon_n$ and 
$|\phi_n(t)\rangle$
are not defined uniquely. Instead, given a solution $\epsilon_n$ and $|\phi_n(t)\rangle$, 
alternative choices are 
given by $\epsilon_{nm}=\epsilon_n+m\omega$ and 
$|\phi_{nm}(t)\rangle=e^{\im m\omega t}|\phi_n(t)\rangle$
with $\omega=2\pi/T$ and $m\in\mathbb Z$, resulting in the same Floquet state
\begin{align}
	|\psi_n(t)\rangle=e^{-\im\epsilon_nt}|\phi_n(t)\rangle=e^{-\im\epsilon_{nm}t}|\phi_{nm}(t)\rangle\ .
	\label{eq:floquet_states_1}
\end{align}
Plugging eq. \eqref{eq:floquet_states_1} into the Schr\"odinger equation \eqref{eq:seq} yields
the Floquet equation
\begin{align}
	\left(H(t)-\im\frac{d}{dt}\right)|\phi_{nm}(t)\rangle&=\epsilon_{nm}|\phi_{nm}(t)\rangle\ ,
	\label{eq:floquet_eq}
\end{align}
which determines the Floquet modes and quasi-energies.

%The time evolution operator can be expressed
%in terms of Floquet states as
%\begin{align}
%	U(t,t_0)=\sum_ne^{-\im\epsilon_n(t-t_0)}|\phi_n(t)\rangle\langle \phi_n(t_0)|
%\end{align}

The Floquet states are
eigenstates of the time evolution operator over one period, i.e.
\begin{align}
	U(t_0+T,t_0)|\psi_n(t_0)\rangle=e^{-\im\epsilon_nT}|\psi_n(t_0)\rangle\ ,
	\label{eq:floqet_state_eigenstate}
\end{align}
and can therefore be regarded as eigenstates of a \emph{Floquet Hamiltonian} $H_{t_0}^F$ defined by
\begin{align}
	U(t_0+nT,t_0)=e^{-\im H_{t_0}^F nT}\ .
\end{align}
The parameter $t_0$ is an arbitrary gauge choice for the Hamiltonian with the property that
$H_{t_0+T}^F=H_{t_0}^F$. Introducing the corresponding gauge-dependent fast-motion operator 
\begin{align}
	U_{t_0}^F(t)\equiv U(t,t_0)e^{\im H_{t_0}^F(t-t_0)}\ ,
\end{align}
which is time-periodic, $U_{t_0}^F(t+T)=U_{t_0}^F(t)$, the full time evolution operator can be expressed
as
\begin{align}
	U(t_2,t_1)=U_{t_0}^F(t_2)e^{-\im H_{t_0}^F(t_2-t_1)}U_{t_0}^F(t_1)^\dagger\ .
\end{align}
Since the quasi-energies $\epsilon_{nm}$ have no $t_0$-dependence, the family of Floquet Hamiltonians,
$H_{t_0}^F$, is moreover gauge equivalent to an 
\emph{effective Hamiltonian} $H_F$, which has no explicit dependence on the driving phase $t_0$ 
\cite{Goldman}.
The corresponding gauge transformation is determined by a Hermitian \emph{kick operator} $\mathcal K(t)$ 
such that
\begin{align}
	H_F=e^{\im \mathcal K(t_0)}H_{t_0}^Fe^{-\im \mathcal K(t_0)}\ .
\end{align}
Note that in general $H_F$ alone does not generate the dynamics over one period. Nevertheless,
the time evolution operator is still split as
\begin{align}
	U(t_2,t_1)=U_F(t_2)e^{-\im H_F(t_2-t_1)}U_F(t_1)^\dagger\ ,
\end{align}
where the \emph{micromotion operator}
\begin{align}
	U_F(t)=e^{-\im \mathcal K(t)}=U_F(t+T)
	\label{eq:micromotion}
\end{align}
was introduced, and the eigenvalue problem 
\begin{align}
	H_F|u_{nm}\rangle=\epsilon_{nm}|u_{nm}\rangle
\end{align}
determines the Floquet modes
\begin{align}
	|\phi_{nm}(t)\rangle=e^{\im m\omega t}U_F(t)|u_{nm}\rangle\ .
	\label{eq:floquet_modes}
\end{align}

\subsubsection{High frequency expansion of the effective Hamiltonian}
Since they are periodic in time
it is beneficial to view the Floquet modes $|\phi_{nm}(t)\rangle$ as elements 
of the composed \emph{Sambe space}
$\mathcal S=\mathcal H\otimes\mathcal L_T$, where $\mathcal L_T$ is the space of $T$-periodic
square integrable functions \cite{Sambe}. Given $\{|\alpha\rangle\}$ is a basis of $\mathcal H$, 
the vectors
\begin{align}
	|\alpha m\drangle=e^{\im m\omega t}|\alpha\rangle
\end{align}
constitute a basis of $\mathcal S$. Here we introduced the notation $|\cdot\drangle$ for vectors
which are explicitly considered as elements of $\mathcal S$. With the natural scalar product in 
Sambe space we obtain
\begin{align}
	\dlangle\alpha m|\alpha' m'\drangle
	&=\langle\alpha|\alpha'\rangle\frac{1}{T}\int_0^T dt e^{-\im\omega(m-m')t}\nonumber\\
	&=\delta_{\alpha\alpha'}\delta_{mm'}\ .
\end{align}
In these terms the operator
\begin{align}
	Q=H(t)-\im\frac{d}{dt}
\end{align}
acts on $\mathcal S$ and eq. \eqref{eq:floquet_eq} is an eigenvalue problem
\begin{align}
	Q|\phi_{nm}\drangle=\epsilon_{nm}|\phi_{nm}\drangle\ .
	\label{eq:ev_prob}
\end{align}
The matrix elements of $Q$ are
\begin{align}
	\dlangle\alpha'm'|Q|\alpha m\drangle
	=\langle\alpha'|H_{m'-m}|\alpha\rangle+\delta_{m'm}\delta_{\alpha'\alpha}m\omega
	\label{eq:Qels}
\end{align}
with the Fourier components of $H(t)$,
\begin{align}
	H_m=\frac{1}{T}\int_0^Tdte^{-\im m\omega t}H(t)\ .
\end{align}
Eq. \eqref{eq:Qels} reveals the block structure of $Q$ with block indices $m,m'$.
Eckardt and Anisimovas \cite{Eckardt} identified the micromotion operator \eqref{eq:micromotion} as
the operator, which block-diagonalizes \eqref{eq:Qels}, thereby yielding the time-independent
effective Hamiltonian
\begin{align}
	H_F=U_F^\dagger(t)H(t)U_F(t)-\im U_F^\dagger(t)\frac{d}{dt}U_F(t)\ .
\end{align}
Making use of the large separation of diagonal matrix elements for large 
frequencies $\omega$ in eq. \eqref{eq:Qels} they apply degenerate perturbation theory to derive
expansions for the effective Hamiltonian as well as the micromotion operator in powers of $1/\omega$.
As a result they find a way to express the effective Hamiltonian as a series
\begin{align}
	H_F=\sum_{n=0}^\infty\frac{1}{\omega^n}H_F^{(n)}\ ,
\end{align}
which can be used to systematically approximate $H_F$ at high frequencies.
The same holds for the kick-operator, which takes the form
\begin{align}
	\mathcal K(t)=\sum_{n=1}^\infty\frac{1}{\omega^n}\mathcal K^{(n)}(t)\ .
\end{align}
In our analysis we consider contributions to these series up to first order, which are
\begin{align}
	H_F^{(0)}=H_0\ ,\quad
	H_F^{(1)}=\sum_{m=1}^\infty\frac{[H_m,H_{-m}]}{m}
	\label{eq:hfHeff}
\end{align}
and
\begin{align}
	\mathcal K^{(1)}(t)
	=-\im\sum_{m=1}^\infty\frac{e^{\im m\omega t}H_m-e^{-\im m\omega t}H_{-m}}{m}\ .
	\label{eq:hfKick}
\end{align}

\begin{figure}[b]
\includegraphics[]{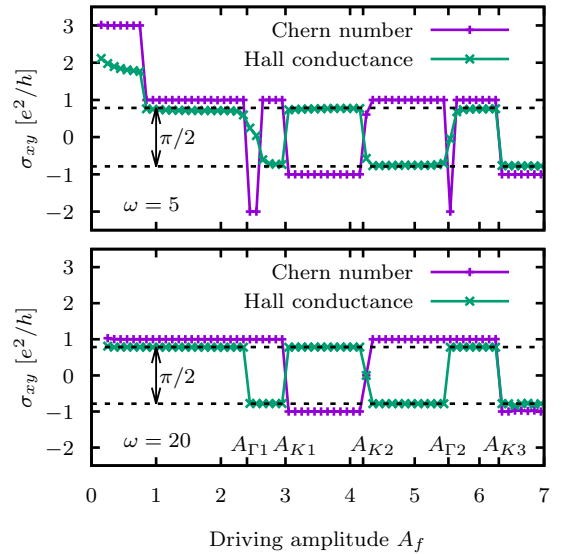}
\caption{Numerical results for the non-equilibrium Hall conductance 
after suddenly switching on the driving with
amplitude $A_f$ with the ground state of the undriven system as initial state for $\omega=5$ (top)
and $\omega=20$ (bottom). The Chern number is computed according to eq. \eqref{eq:tav_chern}.}
\label{fig:hc1}
\end{figure}
\subsection{Non-equilibrium Hall conductance}
In the following we will study the Hall conductance of the stationary state that is reached after a
quench of the driving amplitude at time $t^*$. We assume the system is prepared in an initial state
$|\psi_0(t)\rangle$, which is the ground state of the Hamiltonian with driving amplitude $A_0=0$ 
or a quasi-stationary 
Floquet mode of the driven Hamiltonian with $A_0\neq0$. At time $t_0$ the driving amplitude is suddenly
switched from $A_0$ to $A_1$ and a non-trivial time evolution is induced. We study the Hall conductance
of the state the system reaches a long time after the quench.
Based on linear response theory and using a dephasing argument Dehghani et al. \cite{Dehghani2015} 
derived the following expression for the Hall conductance of a periodically driven electronic
two-band system for this protocol obtaining
\begin{align}
	\sigma_{xy}=\frac{e^2}{2\pi h}\int_{BZ}d^2k\bar F_{\vec kd}
	\left(\rho_{\vec kd}(t^*)-\rho_{\vec ku}(t^*)\right)
	\label{eq:hc_driven}
\end{align}
with the time-averaged Berry curvature
\begin{align}
	\bar F_{\vec kd}&=\frac{2}{T}\int_0^Tdt
	\operatorname{Im}\left[\langle\partial_{k_y}\phi_{\vec kd}(t)|\partial_{k_x}\phi_{\vec kd}(t)\rangle\right]
	\label{eq:tav_bc}
\end{align}
and the occupation numbers of the Floquet modes,
\begin{align}
	\rho_{\vec k\alpha}(t^*)=|\langle\psi_0(t^*)|\phi_{\vec k\alpha}(t^*)\rangle|^2\ .
\end{align}
Here we introduced the indices $\alpha=u,d$ labeling the upper/lower band.
In a cold atom setup with neutral atoms the electron charge $e$ would be replaced by unity.

Dehghani et al. \cite{Dehghani2015} 
considered quenches from the undriven ground state of the graphene Hamiltonian
\eqref{eq:Hamiltonian} to non-zero driving amplitudes $A$. They demonstrated that the Hall 
conductance as a
function of the final driving amplitude changes rapidly whenever the Chern number
\begin{align}
	C=\frac{1}{2\pi}\int_{BZ}d^2k\bar F_{\vec kd}
	\label{eq:tav_chern}
\end{align}
jumps. We reproduced these numerical results as shown in Fig. \ref{fig:scaling_gapped}a and Fig. \ref{fig:hc1} using the method described in appendix \ref{app:numerics} of this paper.

Our results presented in section \ref{sec:results} provide an analytical understanding of the
behavior of the non-equilibrium Hall conductance occurring under this protocol when quenching close
to the transition points.

\subsection{Non-analytic behavior of the Hall conductance of the quenched state for closed systems}
\label{subsec:exp_analysis}
For the case of closed systems Wang et al. \cite{Pei1, Pei2} studied an analogous situation to the
one described above, considering quenches of a parameter $M$ that allows to tune the Hamiltonian
$H(M)$ between different topological phases. In a closed two-band system the Hall conductance
of the stationary state after a quench is
\begin{align}
	\sigma_{xy}=\frac{e^2}{\pi h}\int_{BZ}d^2k
	\operatorname{Im}
	\left[\langle\partial_{k_y} \varphi_{\vec kd}|\partial_{k_x} \varphi_{\vec kd}\rangle\right]
	\left(\rho_{\vec kd}-\rho_{\vec ku}\right)\ ,
	\label{eq:hc_closed}
\end{align}
where $|\varphi_{\vec k\alpha}\rangle$ are the eigenstates of the post-quench Hamiltonian $H(M_f)$ and
$\rho_{\vec k\alpha}=|\langle\psi_0|\varphi_{\vec k\alpha}\rangle|^2$ are the 
occupation numbers of these eigenstates
after the quench. The expressions for the Hall conductance in eq. \eqref{eq:hc_driven} and eq. 
\eqref{eq:hc_closed} have very similar structure and, in fact, also in the case of the closed system
the Hall conductance of the quenched state changes significantly when the quench parameter approaches
an equilibrium phase boundary. In particular, considering the non-equilibrium Hall
conductance close to a phase boundary $M_c$ one finds that
the behavior close to critical points is dominated by the non-analytic part
\begin{align}
	\sigma_{xy}^{\text{div.}}=\frac{e^2}{h}\sum_{\vec q}\mathcal C_\eta^{(\vec q)}(M_i,M_f)
	\label{eq:sigmaeta}
\end{align}
with
\begin{align}\label{eq:hallnonana}
	\mathcal C_\eta^{(\vec{q}_j)} = 
	\int_{\mathcal B_\eta(\vec{q}_j)} \frac{d^2k}{\pi}
	\operatorname{Im}
	\left[\langle\partial_{k_y} \varphi_{\vec kd}|\partial_{k_x} \varphi_{\vec kd}\rangle\right]
	\left(\rho_{\vec kd}-\rho_{\vec ku}\right), 
\end{align}
where $\mathcal B_\eta(\vec{q}_j)$ is a circle of radius $\eta$ centered at
$\vec{q}_j$, the gap-closing points of the quasi-energy spectrum 
in the Brillouin zone. 
If the parameter $M-M_c$ is chosen proportional to 
the gap size the derivative of these contributions diverges as
\begin{align}
	\frac{d\sigma_{xy}^{\text{div.}}}{dM_f}
	\sim\frac{e^2}{h}\frac{C_f^--C_f^+}
	{2|M_i-M_c|}\ln|M_f-M_c|\ ,
	\label{eq:cneqq_closed}
\end{align}
where $C_f^{\pm}$ are the Chern numbers on the right hand side ($+$, $M_f>M_c$) 
and left hand side ($-$) of the
transition, respectively. 
This constitutes a universal non-analytic behavior of the non-equilibrium Hall conductance
$\sigma_{xy}$. The result above is obtained by
expanding the coefficient vector $\vec d_k$, which is for any two-band system defined analogously
to eq. \eqref{eq:Hamiltonian},
around the gap closing points $\vec q$,
\begin{align}
	\vec d_{\vec k}=\vec d_{\vec q}+\hat J_{\vec q}^{\vec d}\Delta\vec k+\mathcal O(\Delta\vec k^2)\ ,
	\label{eq:expansion_closed}
\end{align}
where $\hat J_{\vec q}^{\vec d}$ is the Jacobian matrix of $\vec d_{\vec k}$.
The integral over the Brillouin zone in eq. \eqref{eq:hc_closed} is for $M_f$ close to $M_c$ dominated
by contributions from the vicinity of gap closing points. The remaining part 
$\sigma_{xy}-\sigma_{xy}^\text{div.}$
is an analytic function which is in particular continuous. Any non-analyticity
of the Hall conductance is contributed by $\sigma_{xy}^\text{div.}$.
Note that the terms of $\mathcal O(\Delta\vec k^2)$ do not contribute to the non-analytic behavior,
as discussed in Ref. \cite{Pei2}.
The non-analyticity can therefore be analyzed based on the
expansion to linear order in eq. \eqref{eq:expansion_closed} yielding the result in 
eq. \eqref{eq:cneqq_closed}.

Note that, remarkably, the Dirac cones also lead to universal behavior of the Hall conductance
away from the critical points as shown in Ref. \cite{Unal}.

In the following we will extend this analysis to the case of driven systems based on a high frequency
expansion of the effective Hamiltonian and the micromotion operator.

\section{Results}\label{sec:results}
\subsection{Time-averaged Berry curvature and Berry connection}
\label{subsec:taBC}
An important property of the Berry curvature $\Omega_{\vec k}$ 
in undriven systems is the fact that it can be related to a
local gauge potential, namely the Berry connection $\vec{\mathcal A}_{\vec k}$, via
\begin{align}
	\Omega_{\vec k}=\vec\nabla\times\vec{\mathcal A}_{\vec k}\ .
\end{align}
This property implies through the Kelvin-Stokes theorem that the Chern number is an integer \cite{chen}.

It should be noted that it is a priori not clear whether a corresponding time-averaged Berry connection 
can be attributed to the time-averaged Berry curvature $\bar F_{\vec kd}$ defined in 
eq. \eqref{eq:tav_bc}, because for a non-vanishing
Chern number the Berry connection must exhibit singularities, 
which could prohibit exchanging integrals and derivatives
unheedingly. Nevertheless, we argue in this section that the time-averaged Berry curvature is at least
up to corrections of second order in $\omega^{-1}$ given by the Berry curvature of the effective
Hamiltonian, which is related to a Berry connection.

Applying the product rule for the derivatives the time-averaged Berry curvature 
\eqref{eq:tav_bc} can be 
split into two parts when plugging in eq. \eqref{eq:floquet_modes} for the Floquet modes, yielding
\begin{align}
	\bar F_{\vec kd}&=\Omega_{\vec k}^F\nonumber\\
	&\quad+\frac{2}{T}\int_0^Tdt\operatorname{Im}\Big[
	\langle u_{\vec k}^d|\left(\partial_{k_y}U_F(t)^\dagger\right)
	U_F(t)|\partial_{k_x}u_{\vec k}^d\rangle\nonumber\\
	&\hspace{2cm}
	+\langle\partial_{k_y} u_{\vec k}^d|U_F(t)^\dagger\left(\partial_{k_x}
	U_F(t)\right)|u_{\vec k}^d\rangle\nonumber\\
	&\hspace{1.5cm}
	+\langle u_{\vec k}^d|\left(\partial_{k_y}U_F(t)^\dagger\right)
	\left(\partial_{k_x}U_F(t)\right)|u_{\vec k}^d\rangle
	\Big]
	\label{eq:tav_bc_approx1}
\end{align}
with $\Omega_{\vec k}^F=2
\operatorname{Im}[\langle\partial_{k_y}u_{\vec k}^d|\partial_{k_x}u_{\vec k}^d\rangle]$ the
Berry curvature of the effective Hamiltonian $H_F$. For the derivatives of the operator exponentials
$U_F(t)=\exp(-\im \mathcal K(t))$ we employ the identity
\begin{align}
	\frac{d}{d\lambda}e^{-\im \mathcal K(t)}
	=\int_0^1dse^{-(1-s)\im \mathcal K(t)}\frac{d\mathcal K(t)}{d\lambda}e^{-s\im \mathcal K(t)}
\end{align}
given in Ref. \cite{Wilcox}. This reveals that the last term in eq. \eqref{eq:tav_bc_approx1} is of 
$\mathcal O(\omega^{-2})$, because $\mathcal K(t)\sim\mathcal O(\omega^{-1})$.
For the remaining terms the Baker-Campbell-Hausdorff formula yields
\begin{align}
	\left(\partial_{k_y}U_F(t)^\dagger\right)U_F(t)
	&=\int_0^1dse^{\im(1-s)\mathcal K(t)}\frac{\partial \mathcal K(t)}{\partial k_y}e^{-\im(1-s)\mathcal K(t)}\nonumber\\
	&=\frac{d\mathcal K(t)}{dk_y}+\frac{\im}{2}[\mathcal K(t),\partial_{k_y}\mathcal K(t)]+\ldots\nonumber\\
	&=\frac{d\mathcal K(t)}{dk_y}+\mathcal O(\omega^{-2})
\end{align}
The ellipsis after the second equality stands for higher nested commutators with $K(t)$, which are
all of higher order in $\omega^{-1}$. The analogous argument yields 
$U_F(t)^\dagger\left(\partial_{k_y} U_F(t)\right)=\frac{d\mathcal K(t)}{dk_x}+\mathcal O(\omega^{-2})$. 
Now, according to eq. \eqref{eq:hfKick}, the time dependence 
of the first order contribution to the kick operator, 
$\mathcal K^{(1)}(t)$, is given as a sum of $e^{\im m\omega t}$ 
with $m\neq 0$. Hence, $\int_0^Tdt\partial_{k_{x/y}}\mathcal K^{(1)}(t)=0$ and we obtain
\begin{align}
	\bar F_{\vec kd}=\Omega_{\vec k}^F+\mathcal O(\omega^{-2})\ .
	\label{eq:tav_bc_hf}
\end{align}

Note moreover, that if despite the singularities in $\bar F_{\vec kd}$ the time integral and
derivatives with respect to $\vec k$ can be exchanged the time-averaged Berry curvature can be
written as the curl of a time-averaged Berry connection
\begin{align}
	\bar{\mathcal A}_{\vec kd}^\alpha=
	\frac{1}{T}\int_0^Tdt\langle \phi_{\vec kd}(t)|\partial_{k_\alpha}|\phi_{\vec kd}(t)\rangle
\end{align}
meaning that due to the usual arguments the Chern number 
$C=\frac{1}{2\pi}\int_{BZ}d^2k\vec\nabla \times\bar{\mathcal A}_{\vec kd}$
is an integer. That is, however, only possible if all higher order terms in eq. \eqref{eq:tav_bc_hf} vanish
and $C$ is identically the Chern number of the effective Hamiltonian $H_F$.

\subsection{High frequency expansion}
For the subsequent analysis it is useful to formulate both the high frequency expansion of the
effective Hamiltonian and the expansion of the kick operator in terms of coefficient vectors 
$\vec h_{\vec k}$
and $\vec g_{\vec k}(t)$ such that in the single momentum sectors
\begin{align}
	H_{\vec kF}=\vec h_{\vec k}\cdot\vec\sigma
	\label{eq:heff_coeff}
\end{align}
and
\begin{align}
	\mathcal K_{\vec k}(t)=-\vec g_k(t)\cdot\vec\sigma\ .
\end{align}
In this section we present expressions for the time averaged Berry curvature and the Floquet mode
occupation based on expansions of the respective coefficient vectors.

We will from now on set the hopping $t_h\equiv1$. This means that the high frequency expansion
is valid for $\omega/t_h=\omega\gg1$.

\subsubsection{Effective Hamiltonian and Berry curvature}
\begin{figure}[!h]
\center
\includegraphics[width=.5\textwidth]{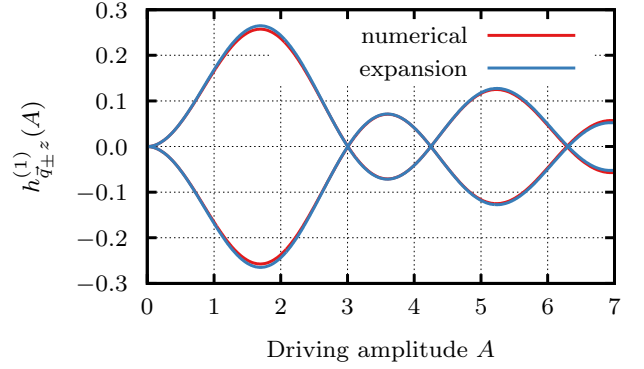}
\caption{Values of the first order contribution to the $z$-component of the coefficient vector at the Dirac
points, $h_{\vec q_\pm z}^{(1)}(A)/\omega$, for $\omega=10$
as function of the driving amplitude $A$ in comparison with the numerical
result for $h_{\vec q_\pm z}(A)$.}
\label{fig:dz1_roots}
\end{figure}

For the high frequency expansion of the effective Hamiltonian given in eq. \eqref{eq:hfHeff} we need the 
Fourier components of the time-dependent Hamiltonian $H_{\vec k}(t)=\vec d_{\vec k}(t)\cdot\vec\sigma$. These
are determined by
\begin{align}
	d_{\vec kx}^m&=\frac{1}{T}\int_0^Tdte^{\im m\omega t}d_{\vec kx}(t)\nonumber\\
	&=-J_m(A)\sum_{j=1}^3\frac{1}{2}e^{-\im m\psi_j}
	\left[e^{\im\vec k\cdot\vec\delta_j}+(-1)^me^{-\im\vec k\cdot\vec\delta_j}\right]\nonumber\\
	d_{\vec ky}^m
	&=-J_m(A)\sum_{j=1}^3\frac{-\im}{2}e^{-\im m\psi_j}
	\left[e^{\im\vec k\cdot\vec\delta_j}+(-1)^{m+1}e^{-\im\vec k\cdot\vec\delta_j}\right]\ ,
	\label{eq:dfourier}
\end{align}
where $J_m(x)$ denotes the $m$-th Bessel function and $\psi_j=\arctan(\delta_j^y/\delta_j^x)$ was
introduced. This yields as the zeroth order term of the effective Hamiltonian just the undriven Hamiltonian
rescaled by the zeroth Bessel function,
\begin{align}
	\vec h_{\vec k}^{(0)}(A)=-J_0(A)\sum_{j=1}^3
	\begin{pmatrix}\cos(\vec k\cdot\vec\delta_j)\\\sin(\vec k\cdot\vec\delta_j)\\0\end{pmatrix}\ .
\end{align}
As the first order term is the commutator of only the Pauli matrices $\sigma_x$ and $\sigma_y$
there is only a contribution to the $z$-component of the coefficient vector, namely
\begin{align}
	h_{\vec kz}^{(1)}(A)=
	4\sum_{n=1}^\infty \frac{J_n(A)^2\sin\left(\frac{2n\pi}{3}\right)}{n}
	\sum_{j=1}^3\sin\left(\vec k\cdot\vec\gamma_j\right)\ ,
	\label{eq:heff1}
\end{align}
where the next-nearest-neighbor vectors
\begin{align}
	\vec\gamma_1=\vec\delta_1-\vec\delta_3\ ,\ \ 
	\vec\gamma_2=\vec\delta_2-\vec\delta_1\ ,\ \ 
	\vec\gamma_3=\vec\delta_3-\vec\delta_2
\end{align}
were introduced. Note that since $J_n(A)^2/n$ decreases with increasing $n$ the infinite sum
in eq. \eqref{eq:heff1} can for practical purposes safely be approximated by a truncation restricted
to the first few terms. Fig. \ref{fig:dz1_roots} shows the analytical result for $h_{\vec qz}^{(1)}(A)$ for $\omega=10$
in comparison with the numerical result at the Dirac points ($K$-points)
\begin{align}
	\vec q_\pm=\begin{pmatrix}0\\\pm\frac{4\pi}{3\sqrt3a}\end{pmatrix}\ .
	\label{eq:gap_closing_points}
\end{align} 
Both show good agreement, in particular in the vicinity of the roots.

The appearance of the n.n.n.-vectors in the effective Hamiltonian reflects the fact that in real space
the first order contribution to the effective Hamiltonian adds a hopping between next-nearest neighbors.
The resulting effective Hamiltonian corresponds to the famous Haldane model where in this case 
the external driving opens a gap in the quasi-energy spectrum leading to a non-vanishing Chern number 
\cite{haldane,Eckardt,shaken_lattice}.

Omitting possible second order contributions to the time-averaged Berry curvature as discussed in section
\ref{subsec:taBC}
the Chern number \eqref{eq:tav_chern} is solely
determined by the effective Hamiltonian $H_F$ and can be expressed in terms of the
coefficient vector $\vec h_{\vec k}$ as
\begin{align}
	C=\int_{BZ}d^2k\frac{ \left( \displaystyle\frac{\partial \vec{h}_{\vec{k}}}{\partial k_x}\times
 \frac{\partial \vec{h}_{\vec{k}}}{\partial k_y}\right) \cdot \vec{h}_{\vec{k}}}
 {4\pi (h_{\vec{k}})^3}
\end{align}
(cf. \cite{chen}).

Note that there are different possibilities for gap closing points in the quasi-energy spectrum of the
effective Hamiltonian, which reads to first order
\begin{align}
	H_{F\vec k}=h_{\vec kx}^{(0)}\sigma^x+h_{\vec ky}^{(0)}\sigma^y
	+\frac{1}{\omega}h_{\vec kz}^{(1)}\sigma^z+\mathcal O\left(\omega^{-2}\right)\ .
\end{align}
Independent of the driving amplitude the zeroth-order terms have roots at the Dirac points $\vec q_\pm$.
Therefore, roots of $h_{\vec q_\pm z}^{(1)}(A)$ 
as function of the driving amplitude mark gap-closing points.
Moreover, $h_{\vec kz}^{(1)}(A)$ has a root at the $\Gamma$-point $\vec k_\Gamma=(0,0)$ 
independent of driving amplitude. This
means that the quasi-energy spectrum closes at this point at roots of $J_0(A)$, because there the
zeroth order terms vanish on the whole Brillouin zone. We marked the transitions that can be attributed to
gap closing points at $K$ or $\Gamma$ points with labels $A_{Ki}$ and $A_{\Gamma i}$, respectively,
in Fig. \ref{fig:hc1}. In the following analysis we will focus on the transitions with gap closing at the 
$K$-points.

\subsubsection{Micromotion operator and occupation numbers}
\begin{figure}[!h]
\includegraphics[width=.5\textwidth]{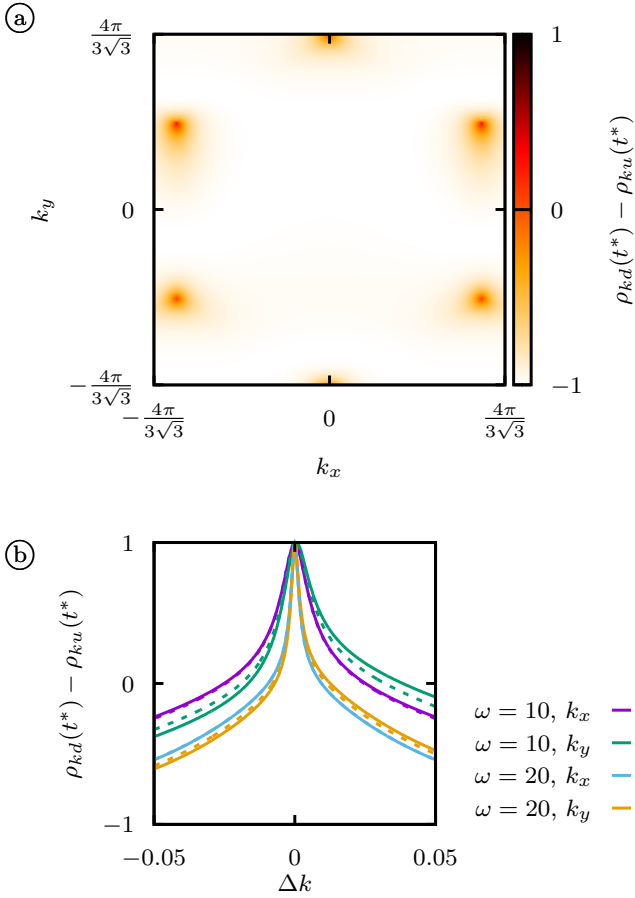}
\caption{a -- Numerical result for the mode occupation after quenching the driving amplitude from
$A_0=0.1$ to $A_1=2.8$ at $t^*=0$. b -- Comparison between numerical results (solid lines) 
and the high frequency expansion \eqref{eq:occ_exp} (dashed lines) 
along cuts with constant $k_x$ or $k_y$, respectively, through the $K$-point $\vec q=(0,-4\pi/3\sqrt3)$
for two different frequencies.}
\label{fig:occ}
\end{figure}

The second ingredient for the Hall conductance of the quenched state is the mode occupation difference
\begin{align}
&\rho_{\vec kd}(t^*)-\rho_{\vec ku}(t^*)\nonumber\\
	&=|\langle\psi_0(t^*)|\phi_{\vec kd}(t^*)\rangle|^2-|\langle\psi_0(t^*)|\phi_{\vec ku}(t^*)\rangle|^2\ ,
	\label{eq:occnums}
\end{align}
which depends on the quench time $t^*$. Fig. \ref{fig:occ}a shows the mode occupation in the
Brillouin zone for a quench from $A_0=0.1$ to $A_1=2.8$ at $t_0=0$. Quenching the amplitude leads to a
smearing of the occupation numbers along the direction of the driving field.
Since we consider quasi-stationary Floquet modes as
initial states, the time dependence is fully determined by the pre- and post-quench micromotion operators.

As given by eq. \eqref{eq:hfKick} the first order term of the high frequency expansion of the kick operator is
\begin{align}
	 \mathcal K_{\vec k}^{(1)}(t)
	=-\im\sum_{m=1}^\infty\frac{1}{m}\left[e^{\im m\omega t}\vec d_{\vec k}^m-e^{-\im m\omega t}\vec d_{\vec k}^m\right]\cdot\vec\sigma
	\label{eq:kick1}
\end{align}
with $\vec d_{\vec k}^m$ given in eq. \eqref{eq:dfourier}.
We approximate the micromotion operator with
\begin{align}
	U_{\vec k}^F(t)=\exp\left(-\im \mathcal K_{\vec k}^{(1)}(t)/\omega\right)+\mathcal O(\omega^{-2})
\end{align}
and define $\vec g_{\vec k}(t)=(g_{\vec kx}(t),g_{\vec ky}(t),g_{\vec kz}(t))$ via
\begin{align}
	- \mathcal K_{\vec k}^{(1)}(t)= \vec g_{\vec k}(t)\cdot\vec\sigma\ .
\end{align}
This approximation of the micromotion operator and the first order result for the eigenvectors of the
effective Hamiltonian yields via eq. \eqref{eq:floquet_modes} the $t_0$-dependent occupation numbers
\begin{align}
	&\rho_{\vec kd}(t^*)-\rho_{\vec ku}(t^*)\nonumber\\
	&=\frac{\vec h_{\vec k}^i\cdot\vec h_{\vec k}^f}{|h_{\vec k}^i||h_{\vec k}^f|}
	+\frac{\left(\vec h_{\vec k}^i\times\vec h_{\vec k}^f\right)\cdot\Delta\vec g(t^*)}
	{|h_{\vec k}^f||h_{\vec k}^i|}
	+\mathcal O(\omega^{-2})\ .
	\label{eq:occ_exp}
\end{align}
Here $\Delta\vec g_{\vec k}(t^*)=\vec g_{\vec k}^{A_f}(t^*)-\vec g_{\vec k}^{A_i}(t^*)$ 
denotes the difference of the Kick operator coefficients before and after switching the driving amplitude.
A detailed derivation of this result is given in appendix \ref{app:occnums}.

Fig. \ref{fig:occ}b shows a comparison between the analytical result in eq. \eqref{eq:occ_exp} 
and the numerical result on cuts through a $K$-point with constant $k_x$ and $k_y$, 
respectively, for
two different driving frequencies. The truncated high frequency expansion clearly 
captures the anisotropy introduced
by the external field and the agreement with numerics improves as the driving frequency is increased.

In order to analyze the non-analytic part of the Hall conductance \eqref{eq:hallnonana} it is crucial
that the occupation numbers contribute a factor $|h_{\vec k}^f|^{-1}$, because thereby the denominator
becomes a polynomial and it is possible to find the antiderivative of the integrand. The result in
eq. \eqref{eq:occ_exp} shows that the correction is proportional to $|h_{\vec k}^f|^{-1}$.
Moreover, the first order contribution to the occupation numbers is an odd function of 
$\Delta\vec k=\vec k-\vec q_\pm$. Therefore, as discussed in Ref. \cite{Pei2},
the corresponding part of the integrand will not contribute
to the non-analytic behavior of the Hall conductance \eqref{eq:tav_bc}. This means that close to
the phase boundaries any dependence of the Hall conductance on the quench time is a second order
contribution in powers of the inverse frequency. We will therefore ignore it in the further analysis.

Experimental setups with finite ramping times will usually not be able to prepare initial states with 
a completely
filled lower band and an empty upper band. Nevertheless, we will focus on this situation in
the following analysis and discuss the effect of partially filled bands as initial states 
later in section \ref{sec:pfb}.

\subsection{Universal behavior at the phase transition}
Putting together eqs. \eqref{eq:tav_bc}, \eqref{eq:tav_bc_hf}, and \eqref{eq:occ_exp}
the non-equilibrium Hall conductance
is determined by
\begin{align}\label{eq:halltwoband}
  \sigma_{xy} &= \frac{e^2}{h}\int d \vec{k}^2 
  \frac{\left(\vec{h}^f_{\vec{k}} \cdot \vec{h}^i_{\vec{k}} 
  \right) \left( \displaystyle\frac{\partial \vec{h}^f_{\vec{k}}}{\partial k_x}\times
 \frac{\partial \vec{h}^f_{\vec{k}}}{\partial k_y}\right) \cdot \vec{h}^f_{\vec{k}}}
 {4\pi {h}^i_{\vec{k}} (h^f_{\vec{k}})^4}
\end{align}
where the integral is over the Brillouin zone. 

We will analyze the non-analytic behavior of \eqref{eq:halltwoband} based on 
expansions of the integrand
around the gap closing points as summarized in section \ref{subsec:exp_analysis} and discussed
more extensively in Ref. \cite{Pei2}. According to the high frequency expansion to first order in powers
of $\omega^{-1}$, we
can suppose the coefficient vectors $\vec h_{\vec k}^{i/f}$ of the initial and final Hamiltonian, respectively,
around some singularity $\vec{q}$ to be
\begin{align}
h^{i/f}_{\vec{k}x} &= J_0(A_{i/f}) \left( a_{1x}\Delta k_x + a_{1y} \Delta k_y\right)+\mathcal O(\Delta k^2) 
\label{eq:exp1}\\
h^{i/f}_{\vec{k}y} &= J_0(A_{i/f}) \left( a_{2x}\Delta k_x + a_{2y} \Delta k_y\right)+\mathcal O(\Delta k^2) \\
h^{i/f}_{\vec{k}z} &= m(A_{i/f})+\mathcal O(\Delta k^2)\ ,
\label{eq:exp3}
\end{align}
where $\Delta k_{x/y}=k_{x/y}-q_{x/y}$.
Note that $A_i$ and $A_f$ are the free parameters and $m(A_{i/f})=h_{\vec qz}^{(1)}(A_{i/f})$ 
is also a function of $A_{i/f}$. In contrast to the closed system analyzed in Refs. \cite{Pei1,Pei2}
for the driven system the expansion coefficients of all components depend on the external parameter,
namely the driving amplitude $A$. However, by introducing the expansion
\begin{align}
	h^{*i/f}_{\vec{k}x} &= a_{1x}\Delta k_x + a_{1y} \Delta k_y+\mathcal O(\Delta k^2)
	\label{eq:modified_exp1} \\
	h^{*i/f}_{\vec{k}y} &= a_{2x}\Delta k_x + a_{2y} \Delta k_y+\mathcal O(\Delta k^2) \\
	h^{*i/f}_{\vec{k}z} &= \frac{m(A_{i/f})}{J_0(A_{i/f})}
	+\mathcal O(\Delta k^2)\equiv \tilde m_{i/f}+\mathcal O(\Delta k^2).
	\label{eq:modified_exp3}
\end{align}
the structure of the integrand in eq. \eqref{eq:halltwoband} remains the same and we obtain the
non-analytic contributions defined in eq. \eqref{eq:hallnonana}, which read
\begin{align}\label{eq:hallnonana1}
\mathcal C_{\eta}^{(\vec{q})}&=\textbf{sgn}(J_0(A_i)) \nonumber\\
&\quad\times\int_{\mathcal B_\eta(\vec{q}_j)} d \vec{k}^2 
  \frac{\left(\vec{h}^{*f}_{\vec{k}} \cdot \vec{h}^{*i}_{\vec{k}} 
  \right) \left( \displaystyle\frac{\partial \vec{h}^{*f}_{\vec{k}}}{\partial k_x}\times
 \frac{\partial \vec{h}^{*f}_{\vec{k}}}{\partial k_y}\right) \cdot \vec{h}^{*f}_{\vec{k}}}
 {4\pi {h}^{*i}_{\vec{k}} (h^{*f}_{\vec{k}})^4}\ .
\end{align}
Therefore, the analysis of the behavior of the Hall conductance close to a transition can be done based
on the expansion \eqref{eq:modified_exp1}-\eqref{eq:modified_exp3} 
with constant coefficients in the first two components given
that $J_0(A_{i/f})\neq0$. As mentioned above, it is sufficient to consider the expansion to linear order. 
In the vicinity of roots of $J_0(A_{i/f})$, however, 
the corresponding expansions of the
first two components of the coefficient vectors are potentially dominated by higher order contributions.

\subsubsection{Quenching from the undriven initial state}\label{subsubsec:undriven}
We first focus on the quenches with the ground state of the undriven system as initial state, i.e. $A_i=0$ 
and $A_f\neq0$.
The gaplessness of the initial Hamiltonian is reflected by $\tilde m_i=0$ in eq. \eqref{eq:modified_exp3}, 
whereas $\tilde m_f\neq 0$. The linear transformation of coordinates
\begin{align}
	\begin{pmatrix}\Delta k_x'\\\Delta k_y'\end{pmatrix}
	=\begin{pmatrix}
	a_{1x}&a_{1y}\\a_{2x}&a_{2y}
	\end{pmatrix}
	\begin{pmatrix}\Delta k_x\\\Delta k_y\end{pmatrix}
\end{align}
allows to make use of the rotational symmetry around the singularity, yielding
\begin{align}
	\mathcal C_{\eta}^{(\vec{q})}&=
	\textbf{sgn}(J_0(A_i))
	\frac{\tilde m_f \textbf{sgn}\left(a_{1x}a_{2y}-a_{2x}a_{1y}\right)}{2} \nonumber\\
	&\quad\times\int^\eta_0 d\Delta k' \frac{\Delta k'^2}{\left(\tilde m_f^2+\Delta k'^2\right)^2}.
\end{align}
The non-vanishing part of this integral in the limit $m_f\to0$ is
\begin{align}
& \frac{\tilde m_f \textbf{sgn}\left(a_{1x}a_{2y}-a_{2x}a_{1y}\right)}
 {4} \int^\eta_0 d\Delta k' \frac{1}{\tilde m_f^2+\Delta k'^2}.\\
 & = \frac{\textbf{sgn}\left(a_{1x}a_{2y}-a_{2x}a_{1y}\right)}
 {4} \arctan \left(\eta/\tilde m_f\right).
\end{align}
For arbitrary $\eta >0$, we have
$\lim_{m_f\to 0}\arctan \left(\eta/\tilde m_f\right)= \textbf{sgn}(\tilde m_f)\pi/2$.
Thus, the discontinuity of $\mathcal C_{\eta}^{(\vec{q})}$ at $m_f=0$ must be
\begin{align}
  &\mathcal C_{\eta}^{(\vec{q})}(m_f\to 0^+) - \mathcal C_{\eta}^{(\vec{q})}(m_f\to 0^-)\nonumber\\
  &= \frac{\pi}{4}\textbf{sgn}\left[J_0(A_i)J_0(A_f)\left(a_{1x}a_{2y}-a_{2x}a_{1y}\right)\right]. 
\end{align}

Summing up the contributions of both $K$-points according to eq. \eqref{eq:sigmaeta} 
yields the discontinuity of the total Hall conductance,
which is
\begin{align}
&  \sigma_{xy}(A_f-A_c\to0^+) - \sigma_{xy} (A_f-A_c\to 0^-) 
\nonumber\\& = \frac{\pi}{4}\textbf{sgn}\left(\frac{J_0(A_i)}{J_0(A_f)}\right)\left[\displaystyle\lim_{m_f\to0^+}C-\displaystyle\lim_{m_f\to0^-}C\right]
\nonumber\\& = \pm\frac{\pi e^2}{2h}. 
\label{eq:jumpheight}
\end{align}
The sign depends on the particular choice of the gap closing point $A_c$.
For critical points which are
related to a closing of the gap at the $K$-points in the Brillouin zone the comparison with 
the numerical results in Fig. \ref{fig:scaling_gapped}a and Fig. \ref{fig:hc1} shows 
that the dimensionless Hall conductance indeed
jumps by $\pi/2$.

\subsubsection{Quenching from a driven initial state}
We now turn to the case where the system is initially prepared in a quasi-stationary Floquet mode
of the driven Hamiltonian with $A_i\neq0$.
When the initial state is a Floquet mode of the driven Hamiltonian, the analysis is completely 
analogous to the case of the closed system in Refs. \cite{Pei1,Pei2}. 
Plugging eqs. \eqref{eq:modified_exp1}-\eqref{eq:modified_exp3} into into eq. \eqref{eq:hallnonana1}
yields the non-analytic part of the integral, which is
\begin{align}
	\mathcal C_{\eta}^{(\vec q)} &\sim 
	-\frac{J_0(A_i)}{J_0(A_f)} 
	\frac{\textbf{sgn}\left(a_{1x}a_{2y}-a_{2x}a_{1y}\right)}  {2\left| m_i\right|}\nonumber\\ 
	&\quad\quad\times m_f(A_f) \ln \left|
 \frac{m_f(A_f)}{J_0({A_f})} \right|\ .
\end{align}
Note first of all that by contrast to quenching from the undriven ground state the Hall conductance is
continuous at the transition points if the initial state is a Floquet mode of the driven Hamiltonian, which
is also evident in Fig. \ref{fig:scaling_gapped}a.
Nevertheless, the derivative with respect to $A_f$ in the limit $A_f\to A_f^c$ is non-analytic
and diverges like
\begin{align}
	\frac{d\mathcal C_{\eta}^{(\vec q)}}{dA_f} \sim & 
	-\frac{J_0(A_i)}{J_0(A^c_f)} 
	\frac{\textbf{sgn}\left(a_{1x}a_{2y}-a_{2x}a_{1y}\right)}{2\left| m_i\right|}
	\nonumber\\ 
	&\quad\quad\times\left.\frac{dm_f}{dA_f}\right|_{A_f=A_f^c} \ln \left|m_f(A_f) \right|.
	\label{eq:gapped_scaling}
\end{align}
Summing up the contributions from both gap closing points in the Brillouin zone yields the divergent
part of the derivative of the Hall conductance
\begin{align}
	\frac{d\mathcal \sigma_{xy}^\text{div.}}{dA_f} \sim & 
	\frac{J_0(A_i)}{J_0(A^c_f)} 
	\frac{\displaystyle\lim_{m_f\to0^+}C-\displaystyle\lim_{m_f\to0^-}C}{2\left| m_i\right|}
	\nonumber\\ 
	& \times\left.\frac{dm_f}{dA_f}\right|_{A_f=A_f^c} \ln \left|m_f(A_f) \right|.
	\label{eq:gapped_scaling1}
\end{align}
Fig. \ref{fig:scaling_gapped}b shows the derivative of the Hall conductance for quenches with different
$A_i$ and $A_f$ close to the transition $K_1$. The derivatives have been rescaled by the respective
prefactors $\mu(A_i,A_f)=J_0(A_f)|m_i|/J_0(A_i)m_f'$ such that according to eq. \eqref{eq:gapped_scaling}
the slopes of all results coincide. Moreover, the results for different $A_i$ have been shifted 
by in order to compare them despite the different regular contributions to the Hall conductance. 
The numerical data agree with the analytically predicted slope and the agreement improves as $A_f$ 
approaches the transition point $A_{K1}$.
Note that after a quench starting with $A_i=2.3$ the agreement is good although this is very close 
to a root of $J_0(A_i)$ as can be seen in the inset of Fig. \ref{fig:scaling_gapped}b.

The presented data were obtained using a grid with 6000$\times$6000 points in the numerical scheme
described in appendix \ref{app:numerics}. This grid resolution determines the computational cost and has
to be increased as $A_f$ approaches $A_{K1}$. Thereby our computational resources limit the
numerical results to the regime presented in Fig. \ref{fig:scaling_gapped}b.

\subsection{The effect of partially filled Floquet bands as initial state}\label{sec:pfb}
As mentioned before the completely filled lower Floquet band we considered above cannot
necessarily be prepared with high fidelity in practice  \cite{dAlessio, Weinberg}. 
In particular, ramping across a gap closing point prohibits adiabatic
preparation of the initial state. Therefore, the initial state will typically 
be given by partially filled Floquet bands
in experiments.

Considering partially filled bands produced using some ramping protocol 
the single particle initial states will be a superposition of the 
pre-quench Floquet modes $|\phi_{0\vec k}^\alpha(t)\rangle$,
\begin{align}
	|\psi_{\vec k}^0(t)\rangle=
	\cos\theta_{\vec k}|\phi_{0\vec k}^d(t)\rangle
	+\sin\theta_{\vec k}e^{\im\varphi_{\vec k}(t)}|\phi_{0\vec k}^u(t)\rangle\ .
	\label{eq:part_occ_ansatz}
\end{align}
In this expression $\theta_{\vec k}$ parametrizes the single particle occupation number and it
will depend on the details of the ramping protocol.
The phase is given by $\varphi_{\vec k}(t)=\varphi_{\vec k}^0+\big(\epsilon_{\vec k}^d-\epsilon_{\vec k}^u\big)t=\varphi_{\vec k}^0-2|\vec h_{\vec k}^{i}|t$, 
where $\vec h_{0\vec k}$ is the coefficient vector of
the initial effective Hamiltonian.
Plugging this into eq. \eqref{eq:occnums} yields
\begin{widetext}
\begin{align}
	\rho_{\vec k}^d(t^*)-\rho_{\vec k}^u(t^*)&=
	\cos(2\theta_{\vec k})
	\Big(|\langle\phi_{0\vec k}^d(t^*)|\phi_{\vec k}^d(t^*)\rangle|^2
	-|\langle\phi_{0\vec k}^d(t^*)|\phi_{\vec k}^u(t^*)\rangle|^2\Big)
	\nonumber\\&\quad
	+\sin(2\theta_{\vec k})\operatorname{Re}\Big[e^{\im\varphi_{\vec k}(t^*)}
	\Big(
	\langle\phi_{0\vec k}^d(t^*)|\phi_{\vec k}^d(t^*)\rangle\langle\phi_{\vec k}^d(t^*)|\phi_{0\vec k}^u(t^*)\rangle
	-
	\langle\phi_{0\vec k}^d(t^*)|\phi_{\vec k}^u(t^*)\rangle\langle\phi_{\vec k}^u(t^*)|\phi_{0\vec k}^u(t^*)\rangle
	\Big)\Big]
	\label{eq:partial_occnum}
\end{align}
\end{widetext}
Thereby, the Hall conductance after a quench can be split into two parts, 
$\sigma_{xy}=\sigma_{xy}^{(1)}+\sigma_{xy}^{(2)}$, corresponding to the first and the second
contribution to the occupation difference.

The first term of the occupation difference above 
equals the occupation difference one obtains when the system is initialized
in the lower Floquet band weighted by the prefactor $\cos(2\theta_{\vec k})$. The specific form 
of the occupation
after preparation, which is parametrized by $\theta_{\vec k}$, will depend on the  
preparation protocol. If $\theta_{\vec k}$ can be approximated by a constant in the vicinity of the
gap closing points $\vec{q}_j$ it will not affect the non-analytic behavior and $\sigma_{xy}^{(1)}$
will contribute a logarithmic divergence to the derivative of the Hall conductance
at the critical time. In the case of the driven hexagonal system considered above the 
non-analyticity in eq. \eqref{eq:gapped_scaling} acquires an additional prefactor
$\cos(\theta_{\vec q})$ with $\vec q$ given in eq. \eqref{eq:gap_closing_points}.

Under the assumption that both $\theta_{\vec k}$ and $\varphi^0_{\vec k}$ are 
well behaved in the vicinity of the gap closing points the second term 
yields a contribution to the non-equilibrium Hall conductance that is independent of the driving
frequency and behaves like the Hall conductance
after quenching from a critical state, as discussed in section \ref{subsubsec:undriven}, 
but is weighted with 
$\sin(\theta_{\vec q})$ and oscillates with frequency $2|\vec h_{\vec q}^{i}|$, i.e. the initial gap width.
For our specific model and the class of critical points we considered above the contribution is
\begin{align}
	\pm\sin(2\theta_{\vec k_0})\cos(\varphi_{\vec k_0}(t^*))\frac{\pi e^2}{2h}
\end{align}
A detailed derivation of this result is given in appendix \ref{app:partfill}. This contribution is non-universal
as it depends on the quench time $t^*$. However, it can in practice be eliminated by averaging over
a range of quench times $t^*$.

Altogether the results obtained for the completely filled lower Floquet band will pertain when 
allowing partially filled Floquet bands as initial
states if the occupation numbers in the vicinity of gap closing points are well behaved. Non-analyticities
in the occupation difference, however, could potentially 
lead to different behavior of the non-equilibrium Hall
conductance.

Any kind of occupation that reflects the spectral properties of a gapped system will be smooth in the
vicinity of the gap closing points $\vec q$. For example, thermal occupation numbers corresponding to an inverse
temperature $\beta$ are obtained from the pure state \eqref{eq:part_occ_ansatz} if 
$\cos\theta_{\vec k}=e^{-\beta\epsilon_{\vec k}^d/2}\big(2\cosh(\beta\epsilon_{\vec k}^d)\big)^{-1/2}$,
where the (quasi-)energies $\epsilon_{\vec k}$ are smooth everywhere. Nevertheless, ramping across
gap closing points could possibly leave an imprint of the non-analyticity in the resulting occupation
numbers. Moreover, it might be possible that the characteristics of the occupation depend on the choice of the ramping protocol.
Such effects, since beyond the scope of this work, should be investigated in the future. 

\section{Discussion}
\subsection{Universality}\label{subsec:universality}
The non-analytic behavior of the Hall conductance at the critical points studied in this work 
is universal in the same sense as discussed in Ref. \cite{Pei2}. 
The key feature
that determines the non-analytic behavior is the conic structure of the quasi-energy spectrum close to
the gap closing point. Thereby, the non-analytic behavior does not depend on the details of the model.

Both expressions characterizing the non-analytic behavior, eq. \eqref{eq:jumpheight} 
and eq. \eqref{eq:gapped_scaling1}, depend only
on the band gap $m(A)$, the band width ratio $J_0(A_i)/J_0(A_f)$, 
and the jump of the Chern number at the transition. 
The Chern number is, however, only defined in translationally invariant systems. Nevertheless, we expect
our results to hold also for systems with weak disorder as we argue in the following. 
Note that this argument regards transitions that occur as a function of the parameter $A$ in the presence of 
weak disorder. Disorder-driven topological transitions at intermediate or strong disorder 
as reported in Refs. \cite{Titum2015, Roy2016} are of different nature and, hence, not in the class of 
transitions we consider in this work.

Other than in undriven topological insulators edge modes of Floquet topological insulators can not only
lie in the energy gap around $\epsilon_{\vec k}=0$. Due to the periodicity of the quasi-energy spectrum
they can also lie in the gap at $\epsilon_{\vec k}=\omega/2=\pi/T$ 
that separates the quasi-energies
of neighboring quasi-energy ``Brillouin zones". The Chern number corresponds to the difference between the 
number of edge modes at $\epsilon=0$, denoted by $\nu_0$, 
and the number of edge modes at 
$\epsilon=\pi/T$, denoted by $\nu_\pi$, i.e. $C=\nu_0-\nu_\pi$. In Ref. \cite{Rudner2013} a bulk invariant
was introduced that directly corresponds to the number of 
edge states in a particular gap for systems with translational invariance. This was generalized to
disordered systems in Ref. \cite{Titum2016}. For the disordered system one adds additional 
time-independent fluxes $\vec\Theta=(\theta_x,\theta_y)$ threaded through the lattice to the
time-periodic Hamiltonian of interest leading to a time evolution operator 
$U(\vec \Theta,t)=\mathcal T_{t'}\exp\big(-\im\int_0^tdt'H(\vec \Theta,t')\big)$.
The number of edge states in a gap around a given quasi-energy $\epsilon$, $\nu_\epsilon$, 
is then determined by
\begin{align}
	\nu_\epsilon=W[U_{\epsilon}]\ ,
	\label{eq:emn}
\end{align}
where
\begin{align}
	W[U_T]&=\frac{1}{8\pi^2}\int_0^Tdt\int d^2\theta
	\nonumber\\&\quad\times
	\operatorname{tr}\Big(
	U_T^{-1}\partial_t U_T\big[U_T^{-1}\partial_{\theta_x}U_T,U_T^{-1}\partial_{\theta_y}U_T\big]
	\Big)
	\label{eq:wn}
\end{align}
is a winding number of the map $U_T(\vec \Theta,t)$ from $(\vec \Theta,t)\in S^1\times S^1\times S^1$
to the space of time evolution operators $U_T(\vec \theta,t)$ periodic in $\theta_x$, $\theta_y$, and $t$.
$U_\epsilon(\vec \Theta,t)$ is related to the time evolution operator of the driven system
$U(\vec \Theta,t)$ via
\begin{align}
	U_\epsilon(\vec k,t)=\Bigg\{
	\begin{matrix}
	U(\vec \Theta,2t)&\text{if }0\leq t\leq T/2\\
	e^{-\im H^\epsilon_\text{eff}(\vec \Theta)t}&\text{if }T/2\leq t\leq T
	\end{matrix}\ .
	\label{eq:return_map}
\end{align}
In this expression $\epsilon$ determines the direction $e^{-\im\epsilon T}$ of the branch cut
of the logarithm in the definition of the effective Hamiltonian
\begin{align}
	H^\epsilon_\text{eff}(\vec \Theta)=\frac{\im}{T}\log_\epsilon U(\vec \Theta,T)\ .
\end{align}

With these results the characteristic non-analytic behavior given by eqs. \eqref{eq:jumpheight} 
and \eqref{eq:gapped_scaling1} can be
reexpressed in terms of the winding number $W$ as
\begin{align}
	\sigma_{xy}^\text{div.}
	\sim
	\frac{\pi}{4}
	\textbf{sgn}\left(\frac{J_0(A_i)}{J_0(A_f)}\right)
	\left[\displaystyle\lim_{m_f\to0^+}\Delta W-\displaystyle\lim_{m_f\to0^-}\Delta W\right]
\end{align}
for quenches from the gapless initial state and as
\begin{align}
	\sigma_{xy}^\text{div.}
	&\sim
	\frac{J_0(A_i)}{J_0(A_f^c)}
	\frac{\displaystyle\lim_{m_f\to0^+}\Delta W-\displaystyle\lim_{m_f\to0^-}\Delta W}{2|m_i|}
	\nonumber\\&\quad\times
	\frac{dm_f}{dA_f}\Big|_{A_f=A_f^c}\ln|m_f|
\end{align}
for the gapped initial state. Here we introduced $\Delta W=W[U_0]-W[U_{\pi/T}]$. 
This form of the non-analytic behavior is expected to pertain also in the presence of weak disorder.
Since beyond the scope of this work it is left for the future to demonstrate this anticipated
behavior explicitly using specific examples.

Note, however, that these results for the non-analytic behavior apply only to transitions with conic 
gap closing points at $\epsilon=0$, which corresponds to points $\vec q$ in the Brillouin zone where
the coefficient vector of the effective Hamiltonian vanishes, $|\vec h_{\vec q}|=0$.
A unique feature of Floquet systems is the possibility of gap closing points at $\epsilon=\pi/T$, which were
for example studied in Refs. \cite{Kitagawa2010,Jiang2011,Kitagawa2012,Rudner2013,Roy2016}. These transitions 
correspond to the presence of points $\vec q$
in the Brillouin zone where $|\vec h_{\vec q}|=\pi/T$. In that case any non-analyticity in the
Hall conductance that is determined by the integral in eq. \eqref{eq:halltwoband} must
originate in non-analytic behavior of the numerator instead of roots of the denominator.
Therefore, our analysis does not apply in these cases.

\subsection{Conclusion}
Based on a high frequency expansion of the effective Hamiltonian and the micromotion operator
we studied the non-equilibrium Hall conductance after sudden switches of the driving amplitude.
Considering a tight binding Hamiltonian on a hexagonal lattice with periodically modulated potential 
we found two kinds of non-analytic behavior
after quenches close to critical driving amplitudes at which the ground state Chern number
exhibits a jump.
%Possible experimental setups are graphene irradiated
%by a circularly polarised laser \cite{driven_graphene_experiment} 
%or ultracold atoms in a periodically shaken optical lattice \cite{shaken_lattice}. 
When the system is initially prepared in the undriven
ground state of the gapless Hamiltonian $H_0$ the non-equilibrium Hall conductance jumps by 
$\pm\frac{\pi e^2}{2h}$
whenever the final driving amplitude $A_f$ crosses a phase boundary $A_c$ of the effective Hamiltonian
with a gap-closing at the $K$-points. Considering neutral atoms in an optical lattice instead of an
electronic system the electron charge $e$ is to be replaced by unity.
If the system is instead initially prepared in a Floquet mode of the driven Hamiltonian $H_{A_i}(t)$ the
non-equilibrium Hall conductance after switching to $A_f$ is continuous at $A_f=A_c$, but the derivative
$\frac{d\sigma_{xy}}{dA_f}$ diverges logarithmically.

This non-analytic behavior is universal in the same sense as discussed in Ref. \cite{Pei2}. 
The characteristics of the non-analyticity only depend on the conic structure of the
quasi-energy spectrum in the vicinity of the gap closing points and are therefore independent of other
details of the model. 
In particular, it is expected that the behavior remains the same in the presence
of weak disorder, where the winding number of the time evolution operator serves as topological
invariant instead of the Chern number.

Nevertheless, at the additional frequency dependent transition points visible in Fig. \ref{fig:hc1}a one
might find different behavior if the gap-closing points have different character. This question should
be addressed in future research.

Our results show that the universal non-analytic behavior of the non-equilibrium Hall conductance
carries over from closed TIs discussed in Refs. \cite{Pei1,Pei2} to FTIs, which can be realized
experimentally in ultracold atom setups in optical lattices 
with the necessary control of external parameters 
\cite{shaken_lattice,Flaschner}. Moreover, small electric fields required to probe the Hall response
can be generated in ultracold atom experiments \cite{Aidelsburger}. These experiments naturally
encounter a situation similar to the one considered in this work, because in the preparation process
the external driving force is usually ramped up at some point in order to bring the system from the initial
topologically trivial state into the topologically non-trivial state of the driven Hamiltonian. It is, however,
understood, that the Chern number of a state is invariant under unitary evolution \cite{dAlessio}. 
In a recent work
\cite{Hu} it was demonstrated how topological properties of the final Hamiltonian can nevertheless be
inferred from the time-averaged non-equilibrium Hall conductance after slow but non-adiabatic ramps.
Our results show that in the opposite limit of infinitely fast ramps the topological invariant
determines the behavior close to transition points. In particular the jump height or the prefactor of the
logarithmic divergence, respectively, are determined by the jump of the 
topological invariant at the transition.
In future work it should be investigated, whether the behavior at infinitely long times investigated here
can be found in the time-averaged Hall conductance at finite times similar to Ref. \cite{Hu}. Moreover, 
the effect of ramping could be studied based on a high frequency expansion as presented in Ref. 
\cite{Anisimovas}.

%% here a revision

%\revision{Insert here the text.
%See fig.~\ref{fig.1}, table~\ref{tab.1} and eq.~(\ref{eq.1}).
%See also~\cite{b.a,b.b}.}

%\begin{equation}
%\label{eq.1}
%0\neq1
%\end{equation}

%\begin{figure}
%\onefigure{epl-template.eps}
%\caption{Figure caption.}
%\label{fig.1}
%\end{figure}

%\begin{table}
%\caption{Table caption.}
%\label{tab.1}
%\begin{center}
%\begin{tabular}{lcr}
%first  & table & row\\
%second & table & row
%\end{tabular}
%\end{center}
%\end{table}

\begin{acknowledgements}
The authors acknowledge helpful discussions with H. Dehghani, S. Kehrein, and D. Huse
and thank L. Cevolani and N. Abeling for proof-reading the manuscript. 
M. Schmitt acknowledges support by the Studienstiftung des Deutschen Volkes. 
P. Wang is supported by NSFC under Grant No. 11304280.
For the numerical computations the Armadillo library \cite{armadillo} was used.
\end{acknowledgements}
%\begin{thebibliography}{0}
%\bibliographystyle{plain}
\bibliography{refs}
%\bibitem{b.a}
%  \Name{Author F., Author S. \and Author T.}
%  \REVIEW{Some Rev. A}{69}{1969}{9691}.

%\bibitem{b.b}
%  \Name{Author F. \and Author S.}
%  \Book{Some Book of Interest}
%  \Editor{A. Editor}
%  \Vol{9}
%  \Publ{Publishing house, City}
%  \Year{1939}
%  \Page{666}.

%\bibitem{b.c}
%  \Editor{Editor A.}
%  \Book{Some Book of Interest}
%  \Vol{9}
%  \Publ{Publishing house, City}
%  \Year{1939}
%  \Section{A}.

%\end{thebibliography}

\clearpage
\newpage
\appendix
\section{Restoring translational invariance by a time-dependent gauge transformation}
\label{app:gauge-trafo}
In order to restore translational invariance we perform a time-dependent gauge transformation with
\begin{align}
	W(t)=\prod_i\exp\left(\im c_i^\dagger c_i\int dt V(\vec r_i,t)\right)
\end{align}
yielding
\begin{align}
	H(t)&=W(t)\tilde H(t)W^\dagger(t)-\im W(t)\partial_tW^\dagger(t)\nonumber\\
	&=-J\sum_{\langle i,j\rangle}\left(e^{-\im\theta_{ij}(t)}c_i^\dagger c_j+h.c.\right)\ ,
\end{align}
where
\begin{align}
	\theta_{ij}(t)=(\vec r_i-\vec r_j)\cdot\vec A(t)
\end{align}
was introduced with
\begin{align}
	\vec A(t)=\frac{V_0a}{\omega}\begin{pmatrix}\sin(\omega t)\\\cos(\omega t)
	\end{pmatrix}\ ,
\end{align}
where $a$ denotes the lattice spacing.

Introducing explicit labels $A$ and $B$ for the sublattices and the Fourier transform of the operators,
\begin{align}
	c_{i,A/B}=\frac{1}{\sqrt N}\sum_{\vec k} e^{-\im \vec k\cdot\vec r_{i,A/B}}c_{\vec k,A/B}
\end{align}
yields the Hamiltonian in momentum space,
\begin{align}
	H(t)=-J\sum_{\vec k} {\vec c_{\vec k}}^{\ \dagger}
	\left[\vec d_{\vec k}(t)\cdot\vec\sigma\right]
	\vec c_{\vec k}\ .
\end{align}
In this expression for the Hamiltonian we introduced
\begin{align}
	\vec c_{\vec k}=\begin{pmatrix}c_{\vec kA}\\c_{\vec kB}\end{pmatrix}
\end{align}
and the coefficient vector $\vec d_{\vec k}(t)$ is given in eqs. \eqref{eq:dcoeff1}-\eqref{eq:dcoeff3} 
in the main text.

\section{Numerical computation of Floquet modes and Hall conductance}\label{app:numerics}
In order to solve eq. \eqref{eq:ev_prob} numerically we set up the matrix $Q$ as given in 
eq. \eqref{eq:Qels} truncating it at some maximal $|m|=M$. The diagonalization of the truncated matrix
yields Floquet modes $|\phi_{nm}\rangle\rangle$ for $-M\leq m\leq M$ and corresponding quasi-energies
with the property $\epsilon_{nm}=\epsilon_{n0}+m\omega$ for small $|m|$. The best approximation
for the eigenvector of the infinite matrix is obtained in the middle of the spectrum, i.e. for $m=0$.

For the two-band system under consideration we obtain a solution at every $\vec k$-point
and the solutions can be written as vectors with $2(2M+1)$ components 
\begin{align}
	\vec \phi_{\vec k,M}^{\alpha m}\equiv|\phi_k^{\alpha m}\rangle\rangle
	=\begin{pmatrix}
	\phi_{k,(u,M)}^{\alpha m}\\\phi_{k,(d,M)}^{\alpha m}\\
	\phi_{k,(u,M-1)}^{\alpha m}\\\vdots\\\phi_{k,(d,-M)}^{\alpha m}
	\end{pmatrix}
\end{align}
The best approximation to the time-dependent Floquet mode $|u_k^\alpha(t)\rangle\in\mathbb C^2$ is
then given by
\begin{align}
	|\phi_k^\alpha(t)\rangle=\sum_{\beta\in\{u,d\}}\sum_{n=-M}^Me^{\im n\omega t}\phi_{k,(\beta,n)}^{\alpha0}|\beta\rangle\ .\label{eq:floquet_mode}
\end{align}
Plugging eq. \eqref{eq:floquet_mode} into eq. \eqref{eq:tav_bc} yields
\begin{widetext}
\begin{align}
	\bar F_{\vec k}&=
	\int_0^T\frac{dt}{T}\sum_{\alpha,\alpha'}\sum_{n,n'}\langle\alpha|e^{-\im n\omega t}
	\partial_x(\phi_{\vec k,(\alpha,n)}^{d0})^*
	\partial_y(\phi_{\vec k,(\alpha',n')}^{d0})e^{\im n'\omega t}|\alpha'\rangle
	\nonumber\\
	&=\sum_{\alpha,\alpha'}\sum_{n,n'}
	\partial_x(\phi_{\vec k,(\alpha,n)}^{d0})^*
	\partial_y(\phi_{\vec k,(\alpha',n')}^{d0})
	\underbrace{\langle\alpha|\alpha'\rangle}_{\delta_{\alpha\alpha'}}
	\underbrace{\int_0^T\frac{dt}{T}e^{\im(n'-n)\omega t}}_{\delta_{nn'}}
	\nonumber\\
	&=\sum_{\alpha}\sum_{n}\partial_x(\phi_{\vec k,(\alpha,n)}^{d0})^*
	\partial_y(\phi_{\vec k,(\alpha,n)}^{d0})
	=\left(\partial_x\vec \phi_{\vec k,M}^{d0}\right)\cdot\left(\partial_y\vec \phi_{\vec k,M}^{d0}\right)	
	\label{eq:av_bc_approx}
\end{align}
\end{widetext}
This means it is not necessary to perform the time-averaging for the averaged Berry curvature explicitly.
The derivatives can be approximated as difference quotients. 
This procedure yields a numerical approximation for $\bar F_k$ on a grid of $\vec k$-points. Choosing
this grid appropriately the Hall conductance \eqref{eq:hc_driven} 
can be computed efficiently using the method
introduced in Ref. \cite{Takahiro} as already established by Dehghani et al. \cite{Dehghani2015}.

\section{Derivation of time dependent mode occupation numbers}
\label{app:occnums}
The first order term of the high frequency expansion of the kick operator is
\begin{align}
	 \mathcal K^{(1)}(t)
	=-\im\sum_{m=1}^\infty\frac{e^{\im m\omega t}H_m-e^{-\im m\omega t}H_{-m}}{m}
	\label{eq:kick1}
\end{align}
Note that in this fomula we dropped the explicit $\vec k$-dependence in the notation, which we will do
also in the rest of this section wherever it is not relevant in order to keep the notation clear.

We approximate the micromotion operator with
\begin{align}
	U_F(t)=\exp\left(-\im \mathcal K^{(1)}(t)/\omega\right)+\mathcal O(\omega^{-2})
\end{align}
and define $\vec g(t)=(g_x(t),g_y(t),g_z(t))$ via
\begin{align}
	- \mathcal K^{(1)}(t)= g_x(t)\sigma^x+ g_y(t)\sigma^y+g_z(t)\sigma^z\ .
\end{align}
Note that according to eq. \eqref{eq:kick1} $g_z(t)=0$ to first order in $1/\omega$.
The expression for $\mathcal K^{(1)}(t)$ as sum of Pauli matrices allows to rewrite the micromotion operator as
\begin{align}
	\exp\left(-\im \mathcal K^{(1)}(t)\right)&=\exp\left[\im g(t) (\vec n(t)\cdot\vec\sigma)\right]
	\nonumber\\
	&=\cos(g(t))+\im \sin(g(t)) (\vec n(t)\cdot\vec\sigma)
\end{align}
with
\begin{align}
	\vec n(t)=\frac{\vec g(t)}{g(t)}\ ,\quad g(t)=|\vec g(t)|\ .
\end{align}
Plugging this expression for the micromotion operator into eq. \eqref{eq:floquet_modes} we obtain 
the overlaps of Floquet modes,
\begin{widetext}
\begin{align}
	\langle\phi_{0}^d(t)|\phi^\alpha(t)\rangle
	&=\langle u_0^d|\left[\cos(g_0(t))-\im\sin(g_0(t))(\vec n_0\cdot\vec\sigma)\right]
	\left[\cos(g(t))+\im\sin(g(t))(\vec n\cdot\vec\sigma)\right]|u^\alpha\rangle\\
	&=\langle u_0^d|u^\alpha\rangle\cos(g_0(t))\cos(g(t))+\sin(g_0(t))\sin(g(t))\langle u_0^d|(\vec n_0\cdot\vec\sigma)(\vec n\cdot\vec\sigma)|u^\alpha\rangle\nonumber\\
	&\quad\quad-\im\sin(g_0(t))\cos(g(t))\langle u_0^d|\vec n_0\cdot\vec\sigma|u^\alpha\rangle 
	+\im\cos(g_0(t))\sin(g(t))\langle u_0^d|\vec n\cdot\vec\sigma| u^\alpha\rangle
\end{align}
\end{widetext}
where the index $0$ indicates Floquet modes/micromotion operator of the initial Hamiltonian with driving amplitude $A_0$.
To evaluate this we need the eigenstates of the effective Hamiltonian
$H_{F\vec k}=\vec h_{\vec k}\cdot\vec\sigma$, which read
\begin{align}
	|u_{\vec k}^{u/d}\rangle
	=\sqrt{\frac{h_{\vec kx}^2+h_{\vec ky}^2}{2|\vec h_{\vec k}|(|\vec h_{\vec k}|\pm h_{\vec kz})}}
	\begin{pmatrix}\frac{h_{\vec kz}\pm|\vec h_{\vec k}|}{h_{\vec kx}+\im h_{\vec ky}}\\
	1\end{pmatrix}\ .
	\label{eq:ev}
\end{align}
These yield the overlaps
\begin{align}
	\Gamma_1^\alpha\equiv\langle u_0^d|u^\alpha\rangle&
	=\frac{1+\frac{h_z^0-|\vec h_0|}{h_x^0-\im h_y^0}\frac{h_z\pm|\vec h|}{h_x+\im h_y}}{\mathcal N_\alpha}\label{eq:Gamma1}\\
	\Gamma_x^\alpha\equiv\langle u_0^d|\sigma^x|u^\alpha\rangle&
	=\frac{\frac{h_z^0-|\vec h_0|}{h_x^0-\im h_y^0}+\frac{h_z\pm|\vec h|}{h_x+\im h_y}}{\mathcal N_\alpha}\\
	\Gamma_y^\alpha\equiv\langle u_0^d|\sigma^y|u^\alpha\rangle&
	=-\im\frac{\frac{h_z^0-|\vec h_0|}{h_x^0-\im h_y^0}-\frac{h_z\pm|\vec h|}{h_x+\im h_y}}{\mathcal N_\alpha}
	\label{eq:Gammay}
\end{align}
where
\begin{align}
	\mathcal N_\alpha^{-1}
	&=
	\sqrt{\frac{(h_x^2+h_y^2)({h_x^0}^2+{h_y^0}^2)}{4|\vec h_0||\vec h|(|\vec h_0|-h_z^0)(|\vec h|\pm h_z)}}
\end{align}
Then
\begin{align}
	&\langle\phi_{0}^d(t)|\phi^\alpha(t)\rangle\nonumber\\
	&=\Gamma_1^\alpha\left[\cos(g_0)\cos(g)+\sin(g_0)\sin(g)(n_0^xn^x+n_0^yn^y)\right]\nonumber\\
	&\quad\quad+\im\Gamma_z^\alpha\sin(g_0)\sin(g)(n_0^xn^y-n_0^yn^x)\nonumber\\
	&\quad\quad-\im\sin(g_0)\cos(g)(n_0^x\Gamma_x^\alpha+n_0^y\Gamma_y^\alpha)\nonumber\\
	&\quad\quad+\im\cos(g_0)\sin(g)(n^x\Gamma_x^\alpha+n^y\Gamma_y^\alpha)\nonumber\\
	&=\Gamma_1^\alpha\left[\cos(g_0)\cos(g)+\sin(g_0)\sin(g)(n_0^xn^x+n_0^yn^y)\right]\nonumber\\
	&\quad+\im\Gamma_y^\alpha(\cos(g_0)\sin(g)n^y-\sin(g_0)\cos(g)n_0^y)\nonumber\\
	&\quad+\im\Gamma_x^\alpha(\cos(g_0)\sin(g)n^x-\sin(g_0)\cos(g)n_0^x)\nonumber\\
	&\quad+\im\Gamma_z^\alpha\sin(g_0)\sin(g)(n_0^xn^y-n_0^yn^x)
\end{align}
Since $g,g_0\sim\mathcal O(\omega^{-1})$, we approximate $\cos(g)\approx 1$ and $\sin(g)\approx g$ 
and drop all terms of $\mathcal O(\omega^{-2})$, which yields
\begin{align}
	&\langle\phi_{0}^d(t)|\phi^\alpha(t)\rangle\nonumber\\
	&=\Gamma_1^\alpha+\im\Gamma_y^\alpha\left(g(t)n^y(t)-g_0(t)n_0^y(t)\right)\nonumber\\
	&\hspace{3em}+\im\Gamma_x^\alpha\left(g(t)n^x(t)-g_0(t)n_0^x(t)\right)\nonumber\\
	&=\Gamma_1^\alpha+\im\Gamma_y^\alpha\underbrace{\left(g_y(t)-g_y^0(t)\right)}_{\equiv \Delta g_y(t)}+\im\Gamma_x^\alpha\underbrace{\left(g_x(t)-g_x^0(t)\right)}_{\equiv \Delta g_x(t)}
\end{align}
Then, again omitting terms quadratic in $1/\omega$,
\begin{align}
	&|\langle\phi_{0}^d(t)|\phi^\alpha(t)\rangle|^2\nonumber\\
	&=(\Gamma_1^\alpha+\im\Gamma_y^\alpha\Delta g_y(t)+\im\Gamma_x^\alpha \Delta g_x(t))
	\nonumber\\
	&\hspace{3em}\times({\Gamma_1^\alpha}^*-\im{\Gamma_y^\alpha}^*\Delta g_y(t)
	-\im{\Gamma_x^\alpha}^* \Delta g_x(t))\nonumber\\
	&=|\Gamma_1^\alpha|^2-\operatorname{Im}[{\Gamma_1^\alpha}^*\Gamma_x^\alpha]\Delta g_x(t)
	-\operatorname{Im}[{\Gamma_1^\alpha}^*\Gamma_y^\alpha]\Delta g_y(t)
\end{align}
In the end we are interested in
\begin{align}
	&|\langle\phi_{0}^d(t)|\phi^d(t)\rangle|^2-|\langle\phi_{0}^d(t)|\phi^u(t)\rangle|^2\nonumber\\
	&=(|\Gamma_1^d|^2-|\Gamma_1^u|^2)\nonumber\\
	&\hspace{1em}-(\operatorname{Im}[{\Gamma_1^d}^*\Gamma_x^d]-\operatorname{Im}[{\Gamma_1^u}^*\Gamma_x^u])\Delta g_x(t)\nonumber\\
	&\hspace{1em}-(\operatorname{Im}[{\Gamma_1^d}^*\Gamma_y^d]-\operatorname{Im}[{\Gamma_1^u}^*\Gamma_y^u])\Delta g_y(t)
\end{align}
The different contributions to the overlaps are
\begin{align}
	|\Gamma_1^d|^2-|\Gamma_1^u|^2&=\frac{\vec d_0\cdot\vec d}{|d||d_0|}\\
	\operatorname{Im}[{\Gamma_1^d}^*\Gamma_x^d]-\operatorname{Im}[{\Gamma_1^u}^*\Gamma_x^u]&=\frac{d_yd_z^0-d_y^0d_z}{|d||d_0|}\\
	\operatorname{Im}[{\Gamma_1^d}^*\Gamma_y^d]-\operatorname{Im}[{\Gamma_1^u}^*\Gamma_y^u]&=\frac{d_x^0d_z-d_xd_z^0}{|d||d_0|}
\end{align}
Since $\Delta g_z(t)=0$, we can finally write
\begin{align}
	&|\langle\phi_{0}^d(t)|\phi^d(t)\rangle|^2-|\langle\phi_{0}^d(t)|\phi^u(t)\rangle|^2\nonumber\\
	&=\frac{\vec h_0\cdot\vec h}{|h||h_0|}+\frac{\left(\vec h_0\times\vec h\right)\cdot\Delta\vec g(t)}{|h||h_0|}
	+\mathcal O(\omega^{-2})\ .
\end{align}

\section{Partially filled bands}
\label{app:partfill}
In this section we derive the leading contribution to
\begin{align}
\sin(2\theta_{\vec k})\operatorname{Re}\Big[e^{\im\varphi_{\vec k}(t)}
	\Big(
	&\langle\phi_{0\vec k}^d(t)|\phi_{\vec k}^d(t)\rangle\langle\phi_{\vec k}^d(t)|\phi_{0\vec k}^u(t)\rangle
	\nonumber\\&
	-
	\langle\phi_{0\vec k}^d(t)|\phi_{\vec k}^u(t)\rangle\langle\phi_{\vec k}^u(t)|\phi_{0\vec k}^u(t)\rangle
	\Big)\Big]
	\label{eq:occnumpart}
\end{align}
which is a part of the occupation difference when quenching from partially filled Floquet bands given
in eq. \eqref{eq:partial_occnum} in the main text. For the sake of brevity we will drop the explicit 
$\vec k$-dependence in the notation wherever it is not relevant in the following.

The derivation is analogous to the one given in appendix \ref{app:occnums}. We generalise the
expressions for the overlaps \eqref{eq:Gamma1}-\eqref{eq:Gammay} to
\begin{align}
	\Gamma_1^{\alpha\beta}=\langle u_0^\alpha|u^\beta\rangle
	&=\frac{1}{\mathcal N_{\alpha\beta}}\Bigg(1+\frac{h_z^0\pm h_0}{h_x^0-\im h_y^0}\frac{h_z\pm h}{h_x+\im h_y}\Bigg)
	\\
	\Gamma_x^{\alpha\beta}=\langle u_0^\alpha|\sigma^x|u^\beta\rangle
	&=\frac{1}{\mathcal N_{\alpha\beta}}\Bigg(\frac{h_z^0\pm h_0}{h_x^0-\im h_y^0}+\frac{h_z\pm h}{h_x+\im h_y}\Bigg)
	\\
	\Gamma_y^{\alpha\beta}=\langle u_0^\alpha|\sigma^y|u^\beta\rangle
	&=-\im\frac{1}{\mathcal N_{\alpha\beta}}\Bigg(\frac{h_z^0\pm h_0}{h_x^0-\im h_y^0}-\frac{h_z\pm h}{h_x+\im h_y}\Bigg)
\end{align}
with
\begin{align}
	\mathcal N_{\alpha\beta}=\sqrt{\frac{\big(h_x^2+h_y^2\big)\big({h_x^0}^2+{h_y^0}^2\big)}{4hh_0\big(h_0\pm h_z^0\big)\big(h\pm h_z\big)}}\ .
\end{align}
In the expressions above $\alpha$ is always associated with the first $\pm$ and $\beta$ with the second.
For the overlaps in eq. \eqref{eq:occnumpart} this yields
\begin{align}
	&\langle\phi_{0\vec k}^d(t)|\phi_{\vec k}^d(t)\rangle\langle\phi_{\vec k}^d(t)|\phi_{0\vec k}^u(t)\rangle
	-
	\langle\phi_{0\vec k}^d(t)|\phi_{\vec k}^u(t)\rangle\langle\phi_{\vec k}^u(t)|\phi_{0\vec k}^u(t)\rangle
	\nonumber\\
	&=
	\Big(\Gamma_1^{dd}+\im\Gamma_x^{dd}\Delta g_x(t)+\im\Gamma_y^{dd}\Delta g_y(t)\Big)
	\nonumber\\&\quad\quad\times
	\Big({\Gamma_1^{ud}}^*-\im{\Gamma_x^{ud}}^*\Delta g_x(t)-\im{\Gamma_y^{ud}}^*\Delta g_y(t)\Big)
	\nonumber\\&\quad-
	\Big(\Gamma_1^{du}+\im\Gamma_x^{du}\Delta g_x(t)+\im\Gamma_y^{du}\Delta g_y(t)\Big)
	\nonumber\\&\quad\quad\quad\times
	\Big({\Gamma_1^{uu}}^*-\im{\Gamma_x^{uu}}^*\Delta g_x(t)-\im{\Gamma_y^{uu}}^*\Delta g_y(t)\Big)
	\nonumber\\
	&=
	\frac{h_z^0\big(h_x^0h_x+h_y^0h_y\big)+\im h_0\big(h_x^0h_y-h_y^0h_x\big)}{hh_0\sqrt{{h_x^0}^2+{h_y^0}^2}}
	\nonumber\\&\quad-
	\frac{4h_z}{hh_0\sqrt{{h_x^0}^2+{h_y^0}^2}}\Big[h_z^0h_y^0+\im h_0h_x^0\Big]\Delta g_x(t)
	\nonumber\\&\quad+
	\frac{4h_z}{hh_0\sqrt{{h_x^0}^2+{h_y^0}^2}}\Big[h_z^0h_x^0-\im h_0h_y^0\Big]\Delta g_y(t)
	+\mathcal O(\omega^{-2})
\end{align}
where second order terms were omitted.

We consider the first term, which is frequency independent.  When the linearisation around the 
gap closing point given in eqs. \eqref{eq:exp1}-\eqref{eq:exp3} is plugged in, the imaginary part 
vanishes. For the real
part we can approximate $h_0\approx h_0^z$ close to the gap closing point. Thereby we obtain
\begin{align}
	\frac{h_x^0h_x+h_y^0h_y}{h\sqrt{{h_x^0}^2+{h_y^0}^2}}
\end{align}
The contribution of this part of the occupation to the Hall conductance is
\begin{align}
	\sigma_{xy}^{(2)}
	&=\int \frac{d^2k}{4\pi} \sin(2\theta_{\vec k})\cos(\varphi_{\vec k}(t))
	\nonumber\\&\quad\quad\quad\times
	\frac{h_{\vec kx}^0h_{\vec kx}+h_{\vec ky}^0h_{\vec ky}}{\sqrt{{h_{\vec kx}^0}^2+{h_{\vec ky}^0}^2}}
	\frac{(\partial_x\vec h_{\vec k}\times\partial_y\vec h_{\vec k})\vec h_{\vec k}}{h_{\vec k}^4}
\end{align}
Assuming $\sin(2\theta_{\vec k})\cos(\varphi_{\vec k}(t))$ well behaved in the vicinity of the gap closing point $\vec k_0$, we can approximate
\begin{align}
	\sigma_{xy}^{(2)}
	&=\sin(2\theta_{\vec k_0})\cos(\varphi_{\vec k_0}(t))
	\nonumber\\&\quad\times
	\int \frac{d^2k}{4\pi}
	\frac{h_{\vec kx}^0h_{\vec kx}+h_{\vec ky}^0h_{\vec ky}}{\sqrt{{h_{\vec kx}^0}^2+{h_{\vec ky}^0}^2}}
	\frac{(\partial_x\vec h_{\vec k}\times\partial_y\vec h_{\vec k})\vec h_{\vec k}}{h_{\vec k}^4}
\end{align}
and then the integral is the same one gets when quenchnig from a gapless initial state, which is
discussed in section \ref{subsubsec:undriven} of the main text. This means, that partially filled 
initial states add a jump
to the Hall conductance at the transition. But through the $\cos(\varphi_{\vec k_0}(t))$ factor the jump oscillates as function of the quench time
with frequency equal to the initial gap $2m(A_i)$.
\end{document}